# Giant tuning of electronic and thermoelectric properties by epitaxial strain in *p*-type Sr-doped LaCrO$_3$ transparent thin films

*Dong Han,*[1] *Rahma Moalla,*[1] *Ignasi Fina,*[2] *Valentina M. Giordano,*[3] *Marc d'Esperonnat,*[1] *Claude Botella,*[1] *Geneviève Grenet,*[1] *Régis Debord,*[3] *Stéphane Pailhès,*[3] *Guillaume Saint-Girons,*[1] *and Romain Bachelet* [1,*]

[1] Univ Lyon, Ecole Centrale de Lyon, INSA Lyon, Université Claude Bernard Lyon 1, CPE Lyon, CNRS, Institut des Nanotechnologies de Lyon - INL, UMR5270, 69134 Ecully, France

[2] Institut de Ciència de Materials de Barcelona (ICMAB - CSIC), Campus UAB, Bellaterra 08193 Barcelona, Spain

[3] Institut Lumière Matière - ILM CNRS UMR 5306, Université de Lyon, UCBL, 10 rue Ada Byron, 69622 Villeurbanne, France

**ABSTRACT**: The impact of epitaxial strain on the structural, electronic, and thermoelectric properties of *p*-type transparent Sr-doped LaCrO₃ thin films has been investigated. For this purpose, high-quality fully-strained La$_{0.75}$Sr$_{0.25}$CrO$_3$ (LSCO) epitaxial thin films were grown by molecular beam epitaxy on three different (pseudo)cubic (001)-oriented perovskite-oxide substrates: LaAlO$_3$, (LaAlO$_3$)$_{0.3}$(Sr$_2$AlTaO$_6$)$_{0.7}$, and DyScO$_3$. The lattice mismatch between the LSCO films and the substrates induces in-plane strain ranging from -2.06% (compressive) to +1.75% (tensile). The electric conductivity can be controlled over two orders of magnitude, $\sigma$ ranging from ~0.5 S cm$^{-1}$ (tensile strain) to ~35 S cm$^{-1}$ (compressive strain). Consistently, the Seebeck coefficient *S* can be finely tuned by a factor of almost two from ~127 µV K$^{-1}$ (compressive strain) to 208 µV K$^{-1}$ (tensile strain). Interestingly, we show that the thermoelectric power factor (PF = $S^2 \sigma$) can consequently be tuned by almost two orders of magnitude. The compressive strain yields a remarkable enhancement by a factor of three for 2% compressive strain with respect to almost relaxed films. These results demonstrate that epitaxial strain is a powerful lever to control the electric properties of LSCO and enhance its thermoelectric properties, which is of high interest for various devices and key applications such as thermal energy harvesters, coolers, transparent conductors, photo-catalyzers and spintronic memories.

## I. INTRODUCTION

Crystalline transition-metal oxides, so-called *functional oxides*, represent an important class of materials since they exhibit a wide range of remarkable properties useful for advanced applications such as non-volatile memories, smart sensors, actuators and energy harvesters. Among others, both thermoelectric oxides (TEO) and transparent conducting oxides (TCO) have attracted rising attention in the last decades because of their potential use as coolers or thermal energy harvesters which can power autonomous wireless sensor networks [1, 2, 3], and/or as transparent conductors in numerous optoelectronic devices [4, 5, 6, 7]. In the broad perovskite-oxide family of general chemical formula $ABO_3$, in which the chemical and the resulting physical properties can be widely tuned, TCO and TEO properties can be adjusted and even both optimized in a single material, as in well-studied titanates such as *n*-type La-doped $SrTiO_3$ that can exhibit both transparency in the visible-infrared range (more than 60%) together with high electrical conductivity and large thermoelectric power factor at room temperature (around 40 µW cm$^{-1}$ K$^{-2}$) [8, 9, 10, 11, 12, 13]. These last few years, advanced *n*-type TCO have been intensively investigated, *e.g.* high-mobility stannates such as La-doped $BaSnO_3$, the electrical conductivity and thermoelectric properties of which can be tuned by aliovalent cationic substitution [14, 15, 16, 17, 18], vanadates such as $SrVO_3$ [4, 19, 20], and more recently molybdates such as $SrMoO_3$ [21, 22]. In contrast, effective *p*-type TCO or transparent TEO are still to be found, which is a crucial issue for the continued development of oxide-based thermoelectric and optoelectronic devices. Sr-doped $LaCrO_3$ ($La_{1-x}Sr_xCrO_3$ solid solutions) has shown for more than forty years both *p*-type conduction and thermoelectricity [23], and more recently rather good optical transparency in the visible and near-infrared range [24]. It has received increasing attention as a *p*-type transparent thermoelectric perovskite oxide these last few years [25, 26, 27]. $La_{2/3}Sr_{1/3}VO_3$ has recently been found to be a *p*-type TCO with properties similar to that of $La_{0.75}Sr_{0.25}CrO_3$. Its thermoelectric properties have not been reported yet [28]. Although of critical importance in the thin-film form, requested for advanced devices such as on-chip thermal energy harvesters or coolers, transparent diodes/electrodes, photocatalyzers or spintronic memories [24, 29, 30, 31], the impact of strain on the physical properties of $La_{1-x}Sr_xCrO_3$ has not been studied yet.

Epitaxial strain is a powerful lever to tune the structural and physical properties of a wide range of materials [32,33, 34]. Strain engineering has been successfully applied to various transition metal perovskite oxides of different properties such as ferroic, conducting, phase changes or metal-insulator transitions (MIT) [33, 34, 35, 36, 37, 38, 39, 40, 41]. Even subtle, strain in transition metal perovskite oxides modifies the fundamental B-O bond length and/or <B-O-B> bond angle through $BO_6$ octahedral distortions, tilts and rotations, and this affects the electronic band structure such as $d$-orbital bandwidth, degeneracy, occupancy, hybridization and ordering, with strong impact on the macroscopic physical properties. More particularly in conducting perovskite oxides, intersite hopping is established from overlap (hybridization) between transition metal B $d$-orbitals and ligand O $p$-orbitals, which depends on strain [36, 37, 40, 42]. For instance, it has been shown that epitaxial strain can be an efficient parameter to tune electrical resistivity and/or metal-insulator transitions in manganites such as $La_{1-x}Sr_xMnO_3$ [36, 43], ruthenates such as $SrRuO_3$ [38, 44, 45, 46], nickelates such as $LaNiO_3$ [40], or vanadates such as $SrVO_3$ [19]. Roughly speaking (*i.e.* disregarding possible consequences of orbital ordering, Jahn-Teller distortion, symmetry lowering and bond disproportionation effects) [47], in-plane tensile strain (or enlargement of unit-cell volume) tends to increase electrical resistivity and/or MIT temperature (mainly because of a decrease orbitals overlap, bandwidth and delocalized states), as in nickelates [40, 48], manganites [36, 43], ruthenates [38, 44, 45, 46], and vanadates [19], and compressive strain (or reduction of unit-cell volume) induces a decrease of electrical resistivity and/or MIT temperature (mainly because of increase of orbitals overlap, bandwidth and delocalized states) as observed in nickelates [40]. Consistently, high hydrostatic pressure studies confirm this tendency in nickelates [40] and even beyond oxides in more conventional thermoelectric materials for which compressive strain in the few GPa range (equivalent of epitaxial strain) tends to enhance the mobile-charge carrier concentration by reducing the bond length and the bandgap [49]. It is worth reminding that thermopower (thermoelectric Seebeck coefficient) and electrical resistivity have similar tendencies [9, 11, 12, 24, 49, 50].

LaCrO$_3$ is a G-type antiferromagnetic insulator (T$_N$ ~ 290 K) with charge transfer gap of 3.3 eV and orthorhombic structure [51, 52, 53]. While the electrical conductivity ($\sigma$) and Seebeck coefficient (*S*) of La$_{1-x}$Sr$_x$CrO$_3$ can be tuned by the partial aliovalent cationic substitution ratio x on the A-site of the perovskite structure, which forms shallow impurity energy level and associated mobile hole carriers [23, 54], the corresponding thermoelectric power factor (PF = $S^2\,\sigma$) being maximum for x around 0.25 keeping rather good optical transparency [23, 24]. At room temperature, La$_{0.75}$Sr$_{0.25}$CrO$_3$ (LSCO) is a transparent thermoelectric semiconductor with a orthorhombic bulk structure (*a* = 5.503 Å, *b* = 5.467 Å, *c* = 7.76 Å, and <Cr-O-Cr> bond angles around 162°, both rather temperature-independent), pseudo-cubic bulk lattice constant of 3.876 Å, and within 25% Cr$^{4+}$ 3*d* $t_{2g}^2$ electronic configuration [24, 27, 55]. No study has been reported yet concerning the strain impact on *p*-type Sr-doped LaCrO$_3$ properties, despite its technological importance as *p*-type TEO and TCO and the strain-engineering potential. However, preliminary observations have mentioned that the epitaxial strain on La$_{1-x}$Sr$_x$CrO$_3$ films can critically impact its electronic transport properties: purely metallic behavior has been observed at x ≥ 0.65 only for compressively strained epitaxial thin films [25]. In addition to partial aliovalent cationic substitution (doping level), the epitaxial strain is expected to be a powerful lever to tailor its physical properties. Indeed, in this paper, we show that epitaxial strain is an efficient parameter to control to a large extent the electronic and thermoelectric properties of *p*-type LSCO transparent thin films, which is of critical importance for integrated LSCO-based devices.

## II. METHODS

About 30 nm thick $La_{0.75}Sr_{0.25}CrO_3$ (LSCO) epitaxially-strained thin films were grown on three single-side polished (001)-oriented (pseudo)cubic perovskite-oxide substrates with different lattice parameters: $LaAlO_3$ (LAO), $(LaAlO_3)_{0.3}(Sr_2AlTaO_6)_{0.7}$ (LSAT), and $DyScO_3$ (DSO) from SurfaceNet GmbH. They provide in-plane strain ($\varepsilon_{xx}$) ranging from -2.06% to +1.75% due to lattice mismatch (see Notes and Figure 1). The samples shown here were also used for another study published elsewhere [27], and were grown successively in the same period and the same growth conditions after proper calibration [27, 56].

The films were grown by solid-source oxide molecular beam epitaxy (MBE) in an ultra-high vacuum (UHV) chamber with a base pressure of less than $1\times10^{-9}$ Torr. Sr, La, and Cr were evaporated from effusion cells in co-deposition. The fluxes were calibrated using a Bayard-Alpert Gauge and checked with a quartz microbalance in UHV (Fig. S1) [27, 56]. The substrates were annealed under the growth conditions for 15 minutes to remove surface impurities before starting the growth at 750 °C and $P(O_2) = 1\times10^{-7}$ Torr, and at a growth rate of 0.15 nm per minute. After deposition, the substrate temperature was lowered to 200 °C at a rate of 50 °C per minute while the background $O_2$ was pumped out. All LSCO films were subjected to post-growth *ex-situ* annealing in air at 300 °C for 2 hours to ensure complete oxidation of the perovskite phase and re-oxidation of the oxide substrates. Backside resistivity measurements were carried out to confirm that the substrates were insulating, which is crucial to avoid artifacts during transport and thermoelectric measurements [57].

Reflection high-energy electron diffraction (RHEED) was used to monitor the epitaxial growth (Fig. 1). The structural properties of the films were investigated by X-ray diffraction (XRD) and X-ray reflectometry (XRR) using a Rigaku Smartlab diffractometer equipped with a high-brilliance rotating anode and a double-crystal Ge(220) monochromator (Figs. 2, S2-S3, S5, Table S1). Atomic force microscopy was used in tapping mode to characterize the film surface (Figs. S4-S5, Table S1). X-ray photoemission spectroscopy (XPS) was used to characterize the main chemical properties of the films (Fig. S6) [27]. Electrical resistivity was determined using the van der Pauw method in the 100-400 K temperature range. For that purpose, Cr/Au electrodes were placed at the vortexes of the films. These electrical transport measurements were performed in a physical property measurement system (PPMS) from Quantum Design (Fig. S7). Seebeck coefficients ($S$) were measured using the differential method $S = \Delta V/\Delta T$ (Figs. S8-S9), where $\Delta V$ is the thermo-electromotive force induced by the thermal gradient $\Delta T$, in a high-temperature set-up, similar to that described elsewhere [58]. $S(T)$ has then been measured just above room temperature, a temperature range of high interest for most of the applications.

## III. RESULTS AND DISCUSSION

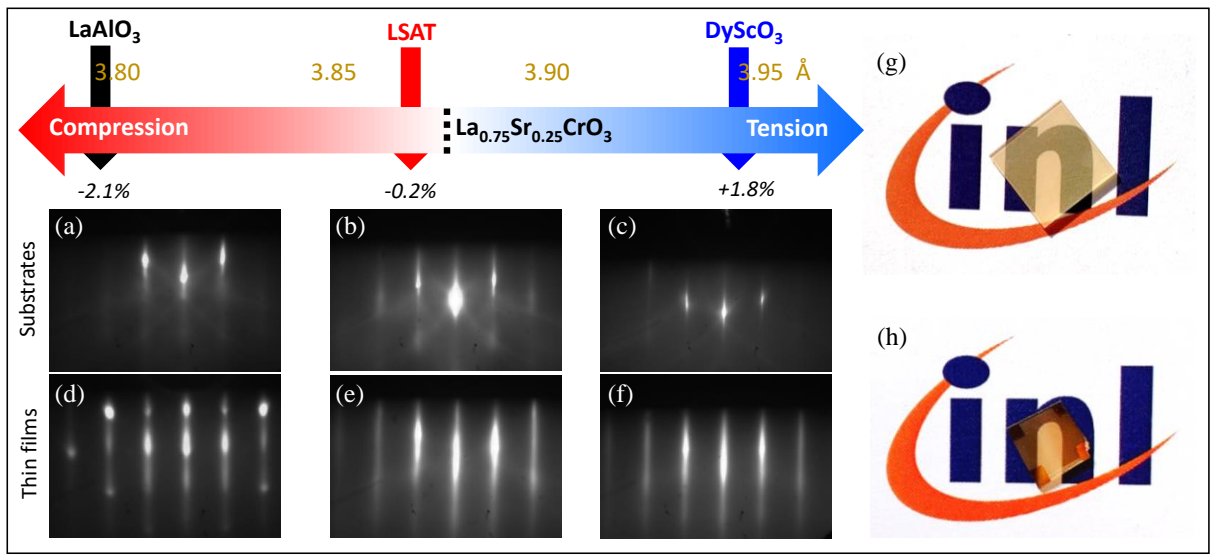

**Figure 1. Lattice mismatch/strain, RHEED patterns, and photographs of the samples.** (a-c) RHEED patterns recorded along the <100> azimuth of the substrates after *in-situ* surface preparation (annealing at 750 °C in $P(O_2)$ = 1×10$^{-7}$ Torr for 15 min). (d-f) RHEED patterns recorded along the <100> azimuth of the LSCO layers at the end of the growth (partly/adapted from Fig. 2 of Ref. 27). Italic figures correspond to the epitaxial strain induced in the LSCO layers by the substrates, and the black vertical dotted line corresponds to 0% strain. (g) Photograph of virgin LSAT substrate and (h) photograph of the 30 nm thick LSCO film grown on LSAT after the deposition of Cr/Au contacts in the corners for electrical characterizations.

According to our former work [27], the lattice constant of bulk LSCO with 25% Sr concentration is 3.876 Å at room temperature, which lies close to the lattice constant of LSAT (Figure 1). LSCO is thus grown almost relaxed on LSAT (exactly under slight compressive strain of -0.21 %), under compressive strain (-2.06 %) on LAO, and under tensile strain (+1.75 %) on DSO (see Notes). The RHEED patterns of the LSCO layers grown on the three substrates (LAO, LSAT, and DSO) are shown in Figure 1(d-f). All the patterns present well-contrasted integer-order reflections attesting for single-phase epitaxial films. For the films on LSAT and DSO, the RHEED reflections are streaky attesting for flat films, whereas for the film with the large compressive strain on LAO (-2.06%), RHEED reflections are slightly spotty, suggesting a certain surface roughness, as frequently observed in strongly compressively-strained epitaxial layers. However, XRR and AFM analysis reveal rather low root-mean-square (rms) surface roughnesses, lower than ~0.7 nm, in the same order of magnitude than the two other samples (see Supporting Information, Figs. S3, S4, S5, and Table S1). For instance, the differences of rms surface roughness between LSCO/LAO and LSCO/LSAT are small (0.14 nm measured macroscopically by XRR, and 0.26 nm measured microscopically by AFM). The LSCO films are confirmed to be rather well transparent (Figure 1g and 1h) with slightly brown color in good agreement with the observations made by Zhang *et al*. (transmittance between ~40 % and ~65 % in the visible and near-infrared range for this LSCO composition and with 80 nm thick films [24]). The LSCO transmittance appears very similar to other emergent perovskite TCO, such as *n*-type $SrVO_3$ for instance (<65% with 45 nm thick films) [4], and somehow lower than standard TCO such as *n*-type Indium-Tin Oxide (ITO) but which is limited by indium scarcity and not of perovskite structure [5, 7].

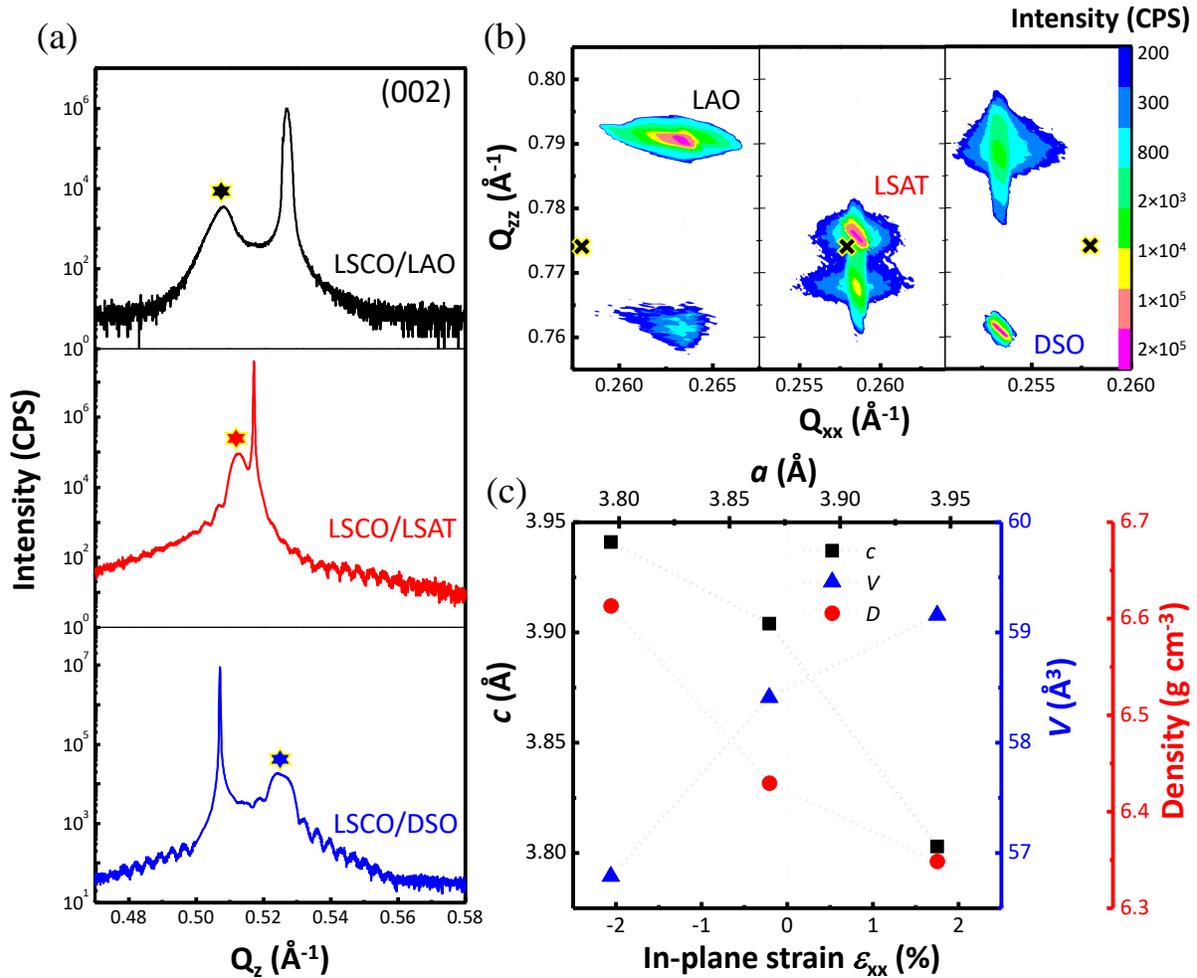

**Figure 2. XRD characterizations.** (a) XRD $2\theta/\omega$ scans (reciprocal space units) around the (002) reflections of the LSCO films and their substrates. The stars indicate the LSCO peak positions. (b) Reciprocal space map recorded around the asymmetric (103) reflection of the films and substrates. All the films are coherently strained on the substrates, as indicated by the grey dashed lines. The black crosses indicate the (103) position of bulk LSCO, taking into account the thermal strain caused by a slight thermal expansion mismatch between the substrates and the films (see Notes and Ref. 59). (c) Out-of-plane lattice parameters ($c$), unit-cell volume ($V$), and density ($D$, extracted from unit cell volume) of LSCO films as a function of the in-plane strain ($\varepsilon_{xx}$). This figure is adapted from Fig. 2 of Ref. 27.

Figure 2 presents the structural properties of all LSCO films investigated by XRD. Wide $2\theta/\omega$ scans, as well as pole figure measurements (not shown here), show LSCO ($00l$) reflections only, indicating that neither other phases than LSCO nor other orientations than ($00l$) are present in the epitaxial films (see Supporting Information, Figs. S2a-c) [27]. Figure 2a shows the $2\theta/\omega$ scans recorded around the (002) reflections of the LSCO films and the substrates. The (002) peaks attributable to the LSCO films are indicated by stars. The presence of Pendellösung fringes around the diffraction peaks of the films, especially on LSAT and DSO substrates, attest to their high crystalline quality and abruptness of their interfaces. These fringes seem to appear more clearly with in-plane tensile strain for which the rms surface roughness is lower (see Supporting Information, Figs. S3, S4, S5, Table S1). All films have low mosaicity, below 0.1° for LSCO on LSAT and DSO, and ~0.2 ° for LSCO on LAO (see Supporting Information, Fig. S2d, Table S1). A gradual shift in the $Q_z$ position of the (002) reflection of the films is observed with in-plane strain in Figure 2a. The corresponding measured out-of-plane lattice parameters ($c_{LSCO}$) of the films for each film ranges from 3.941 Å (in-plane compressive) to 3.804 Å (in-plane tensile), as shown in Figure 2c. The XRD reciprocal space maps (RSM) around the asymmetrical (103) reflection indicates that all LSCO films are coherently strained to their substrates with undetectable strain relaxation (Figure 2b). The in-plane epitaxial strain ($\varepsilon_{xx}$) of the LSCO films varies from -2.06% (compressive) to +1.75% (tensile), depending upon the substrate (see Notes). The out-of-plane strain ($\varepsilon_{zz}$) of the LSCO films ranges from +1.68% (in-plane compressive) to -1.86% (in-plane tensile). Elastic strain is rather large in LSCO because of its relatively large Poisson ratio ($v = 0.32$) [27]. Figure 2c shows that the unit-cell volume of LSCO films ($V_{LSCO}$, blue triangles in Figure 2c) increases from 56.8 Å$^3$ (compressive) to 59.2 Å$^3$ (tensile) [27]. The $V_{LSCO}$ of the film grown almost relaxed on LSAT is about 58.4 Å$^3$, close to the bulk LSCO volume value (58.2 Å$^3$). The density of all LSCO films was then extracted from unit-cell volume and is plotted as a function of strain in Figure 2c (red circles). It decreases quasi-linearly from 6.61 g cm$^{-3}$ (compressive) to 6.35 g cm$^{-3}$ (tensile) ($D_{LSCO\text{-}bulk} = 6.45$ g cm$^{-3}$) with increasing strain.

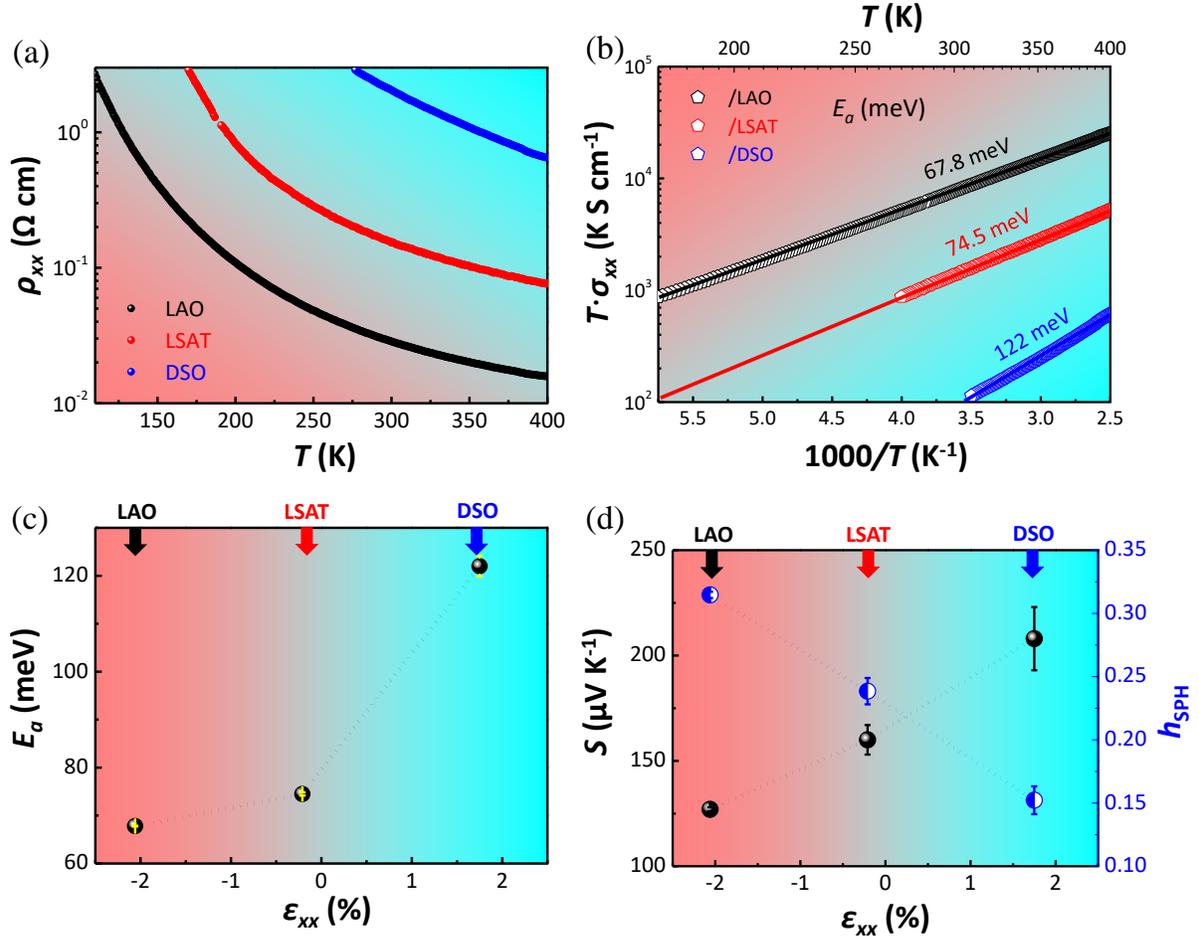

**Figure 3. In-plane electronic and thermoelectric transport characterizations.** (a) Temperature dependence of the resistivity ($\rho_{xx}$) in the 100-400 K range. (b) Corresponding temperature dependence of the conductivity ($\sigma_{xx}$). (c) Activation energies ($E_a$) for small polaron hopping extracted from (b), plotted as a function of in-plane strain $\varepsilon_{xx}$. (d) Room temperature Seebeck coefficient ($S$) and fraction of hopping sites occupied with holes ($h_{SPH}$) estimated from $S$ and small polaron hopping model, plotted as a function of in-plane strain $\varepsilon_{xx}$.

Figure 3 presents the in-plane electronic and thermoelectric transport measurements performed on the LSCO films. First, the temperature dependence of the longitudinal resistivity measured in the range of 100-400 K is shown in Figure 3a. All films behave as semiconductors, as expected with this LSCO composition [23, 25]. However, the resistivity values are very different depending on the film (strain), and they vary over almost two orders of magnitude, the in-plane compressively (tensely) strained film being more conducting (insulating). For instance, the resistivity at 300 K is about ~$10^{-2}$ Ω cm on LAO (compressive strain), ~$10^{-1}$ Ω cm on LSAT (almost relaxed), and ~1 Ω cm on DSO (tensile strain). The compressive LSCO film on LAO has resistivity values that are more comparable to 50% than 25% Sr-doped LCO films, whereas the tensile LSCO film on DSO has resistivity values comparable to 0.04% Sr-doped LCO films [24, 25]. It is worth noting that the almost relaxed LSCO film on LSAT has resistivity values in good agreement with other 25% Sr-doped LaCrO$_3$ films of comparable strain [25].

Figure 3b presents the corresponding temperature dependence of the conductivity, consistent with thermally activated hopping transport in agreement with previous studies [23, 24]. The different slopes indicate different activation energies ($E_a$) for hopping transport that depends upon the strain. In agreement with the resistivity/conductivity trend, $E_a$ increases with in-plane tensile strain and decreases with compressive strain, which is certainly linked with strain-dependent unit-cell volume and Cr 3$d$ - O 2$p$ orbital hybridizations, as in many other complex oxides (see Introduction). $E_a$ for the most conductive film (compressive on LAO) is around 67.8 meV, more comparable to 50% Sr-doped than 25% Sr-doped LCO films, whereas that of the most resistive film (tensile on DSO) is around 122 meV, more comparable to ~4% Sr-doped LCO films [24]. Figure 3c summarizes the $E_a$ values of LSCO plotted as a function of in-plane strain.

Figure 3d presents both the Seebeck coefficients $S$ (measured in-plane, raw data in Supporting Information, Fig. S8) and the fraction of hopping sites occupied with holes ($h_{SPH}$, estimated from $S$ using a small polaron hopping model, see below) as a function of in-plane strain and corresponding stress. The positive $S$ for all samples confirms that holes are the dominant mobile charge carriers in these films, as expected [23, 24, 25]. The Seebeck coefficient monotonously increases with strain (from compressive to tensile) in agreement with the decrease of the electrical conductivity [24, 50], by almost a factor two from 127 µV K$^{-1}$ (compressive) to 208 µV K$^{-1}$ (tensile), consistently with other observations made on various thermoelectric materials under high-pressures (S decreasing with pressure) [49]. It is worth noting that the Seebeck coefficients of the almost relaxed LSCO film on LSAT (160 µV K$^{-1}$) and another moderately tensile on STO substrate (180 µV K$^{-1}$, see Supporting Information, Fig. S9) are in perfect agreement with Zhang et al. [24].

As in Ref. 24, Hall measurements being unsuccessful on these samples most probably because of both low carrier mobility and relatively high resistivity [60], the fraction of hopping sites occupied with holes ($h_{SPH}$) has been estimated from S values using a small polaron hopping model, such as:

$$S = \frac{k_B}{e} \ln\left[\frac{2(1-h_{SPH})}{h_{SPH}}\right]$$

where $k_B$ is Boltzmann's constant, $e$ is the electronic charge [24]. $h_{SPH}$ should be 0.25 for this LSCO composition in bulk (relaxed state). However, it rather strongly deviates here depending on strain: monotonously decreasing from 0.315 (compressive) to 0.238 (almost relaxed) and 0.152 (tensile), consistently with resistivity/conductivity data (Fig. 3a-c). It is worth noting that $h_{SPH}$ is about 0.2 for LSCO grown on SrTiO$_3$ substrate (in slight tensile strain), in very good agreement with this trend (see Supporting Information, Fig. S10) [24]. The mobile hole density ($p_{SPH}$) as well as hole mobility [$\mu_{SPH} = \sigma_{xx} / (e\, p_{SPH})$] have then been estimated from $h_{SPH}$ values (Fig. 4a).

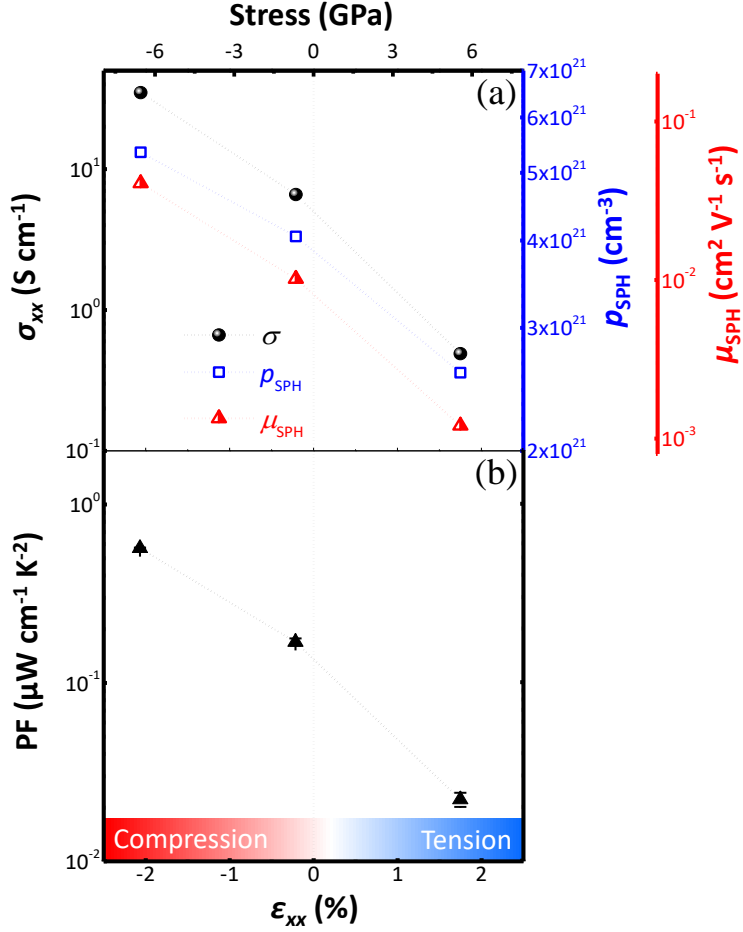

**Figure 4. In-plane electronic transport and thermoelectric properties at 300 K**, plotted as a function of in-plane strain ($\varepsilon_{xx}$) and corresponding stress (see Notes and Refs. 61, 62). (a) Electrical conductivity ($\sigma_{xx}$), as well as hole carrier density ($p_{SPH}$) and mobility ($\mu_{SPH}$) estimated from $S$ using a small polaron hopping model (see main text). (b) Thermoelectric power factor (PF = $S^2\sigma$).

Figure 4 presents the electronic transport and thermoelectric properties of the LSCO films at room temperature as a function of in-plane strain and corresponding stress (see Notes)[61, 62]. Figure 4a presents the values of in-plane electrical conductivity, mobile charge carrier density, and mobility assuming only one-type (hole) carrier. As presented above with resistivity ($\rho_{xx}$), the electrical conductivity ($\sigma_{xx} = \rho_{xx}^{-1}$) at 300 K monotonously decreases by two orders of magnitude from ~35 S cm$^{-1}$ (compressive) to 0.5 S cm$^{-1}$ (tensile), with intermediate ~7 S cm$^{-1}$ (almost relaxed) in good agreement with literature [25]. The extracted mobile hole density ($p_{SPH}$) monotonously decreases from 5.5 10$^{21}$ cm$^{-3}$ (compressive) to 2.6 10$^{21}$ cm$^{-3}$ (tensile), with intermediate value about 4.1 10$^{21}$ cm$^{-3}$ (almost relaxed) in good agreement with it is expected from bulk value at this LSCO composition (25% holes per unit cell leading to $p$ ~4.3 10$^{21}$ cm$^{-3}$). The extracted hole mobility ($\mu_{SPH}$) largely decreases by more than one order of magnitude from ~0.04 cm$^2$ V$^{-1}$ s$^{-1}$ (compressive) to ~0.001 cm$^2$ V$^{-1}$ s$^{-1}$ (tensile), with intermediate value about ~0.01 cm$^2$ V$^{-1}$ s$^{-1}$ (almost relaxed), in rather good agreement with ref. [24]. For general comparisons, hole mobility in LSCO appears to be about one-two orders of magnitude lower than *n*-type La-doped STO epitaxial films [11, 63], and three-four orders of magnitude lower than high-mobility *n*-type La-doped BaSnO$_3$ epitaxial films [16, 18, 64] or *p*-type doped GaAs-based III-V semiconductors (see Notes). It is worth noting that hole mobility is often lower, up to two orders of magnitude, than electron mobility for the same core semiconductor materials, as in ZnO [65, 66] or GaAs (see Notes), for instance.

Figure 4b presents the thermoelectric power factor (PF = $S^2 \sigma$) at room temperature as function of in-plane strain. Interestingly, despite the Seebeck coefficient trend, the thermoelectric power factor is dominated by the electrical conductivity, decreasing largely by more than one order of magnitude from ~0.56 µW cm$^{-1}$ K$^{-2}$ (compressive) to 0.02 µW cm$^{-1}$ K$^{-2}$ (tensile). All these coherent results show that the electronic and thermoelectric properties of LSCO can be finely tuned by lattice strain, and that compressive strain enhances both the electrical conductivity and thermoelectric power factor, which is of high interest for various applications.

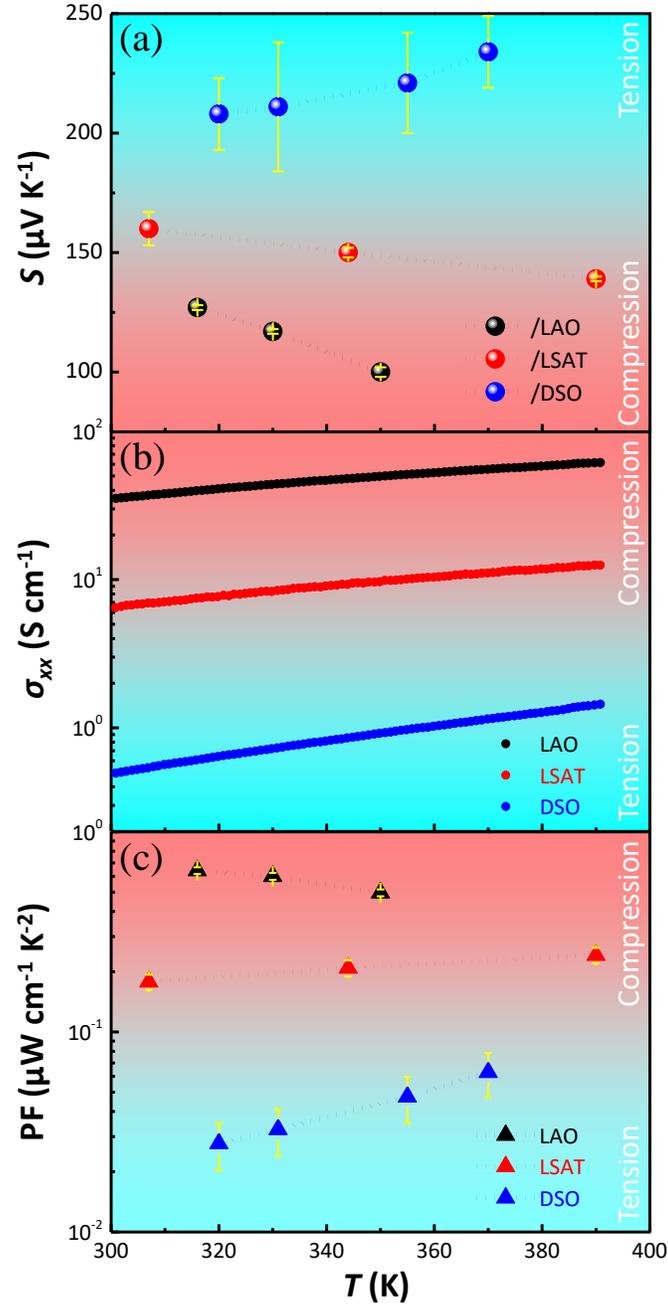

**Figure 5. Temperature-dependent thermoelectric properties.** (a) Seebeck coefficient ($S$), (b) electrical conductivity ($\sigma_{xx}$) and (c) thermoelectric power factor (PF = $S^2 \sigma_{xx}$) for each film as a function of the temperature in the range of 300-400 K.

Figure 5 presents the temperature-dependence of the Seebeck coefficient ($S$), the electrical conductivity ($\sigma_{xx}$) and the thermoelectric power factor (PF) for all LSCO films measured just above room temperature (300-400 K), in a range of high interest for the main applications. The Seebeck coefficients decrease with increasing temperature for the two most conductive samples, driven by the temperature dependence of electrical conductivity. This decrease is all the faster that the film is conductive. In contrast, for the most resistive (tensile) film, $S$ increases with increasing temperature, in contradiction with the temperature dependence of electrical conductivity. However, the $S$ measurement uncertainties are larger for this resistive sample. The $S$ can be considered almost constant with temperature in this range within these uncertainties, which can be observed elsewhere in bulk LSCO [23]. We note that the thermal expansion coefficients of the substrates are very similar (see Notes) so that the slight differences in thermal expansion mismatch between the substrates and the films are too small (and not consistent with the trend) to explain the differences in the observed temperature-dependent variations. The electrical conductivity differences between films are maintained whatever the temperature. Interestingly, considering the measurement uncertainties of $S$, the PF of this resistive film increases slightly with temperature, contrary to the most conductive film (Figure 5c). The PF of the intermediate almost relaxed film appears monotonous with temperature. Nevertheless, no important change is observed with temperature in this range, in agreement with previous studies on structural, electronic and thermoelectric properties of bulk LSCO [23, 55]. At 350 K for instance, depending on the strain, the Seebeck coefficients still differ by a factor of two (Figure 5a), the electrical conductivity by two orders of magnitude (Figure 5b) and the PF by one order of magnitude (Figure 5c), similarly to 300 K.

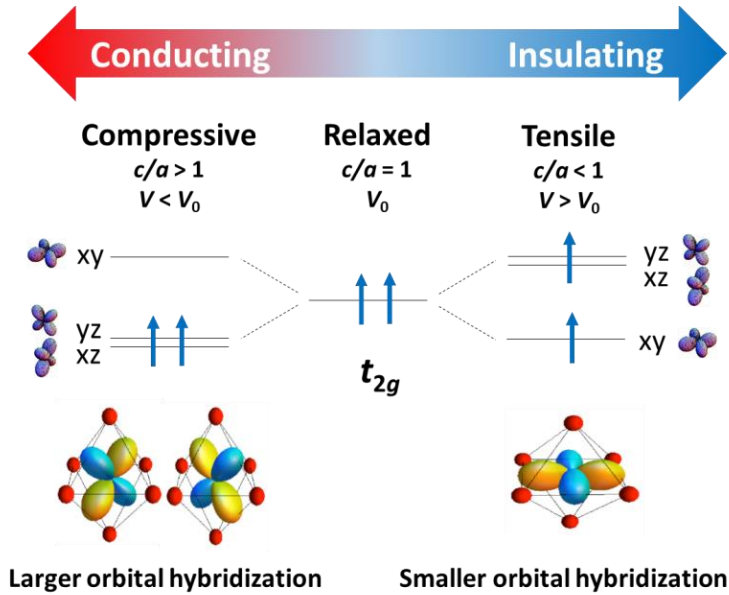

**Figure 6. Summarizing sketch** of the strain-dependent electrical resistivity trend with possible preferential occupancy of the $Cr^{4+}$ $3d$ $t_{2g}^2$ orbitals within the crystal field.

As mentioned in the introduction, the lattice strain of perovskite oxides can largely modify their physical properties, which has led to strain engineering of various thin films by heteroepitaxy [33, 34, 35, 37]. In conducting perovskite oxides ($ABO_3$) with $d$-electron transition metal (B), the lattice strain/distortion (change in the B-O bond length and/or <B-O-B> bond angle) affects the band structure due to strong electron-orbital-lattice couplings, such as bandwidth, orbital hybridizations, orbital level splitting (splitting of $t_{2g}$ and $e_g$ levels depending on the crystalline symmetry), and sub-band filling responsible for possible doping and metal-insulator transitions [37, 42]. In tetragonal LSCO epitaxial films, the Cr $3d$ - O $2p$ orbital hybridization and level at the top of the valence band, together with the empty split Cr $3d$ $t_{2g}$ orbital level at the bottom of the conduction band, which allow hole hopping transport [24], are certainly altered because of such lattice strain/distortion even at constant band filling [37, 42]. As observed with other perovskite oxides such as nickelates [40], the biaxial in-plane tensile strain in LSCO (larger Cr-O bond length, <Cr-O-Cr> bond angle up to 180° and unit cell volume) would decrease the Cr $3d$ - O $2p$ orbital hybridization (bandwidth) at the valence band. This would then decrease the hopping transport probability and consequently decrease the mobile hole density responsible for $p$-type electrical (semi)conductivity by small polaron hopping. Additionally, a Jahn-Teller like distortion may occur as observed in $CrO_2$ [67] or other perovskite oxides such as $La_{2/3}Sr_{1/3}MnO_3$ [39, 68], $LaCoO_3$ [41] and $SrVO_3$ [19] where in-plane tensile strain would lower and favor the occupied in-plane Cr $3d$ $t_{2g}$ (xy) orbital ground state (Figure 6). On the contrary, the biaxial in-plane compressive strain (smaller Cr-O bond length, <Cr-O-Cr> bond angle than 162°, and unit-cell volume) would increase the Cr $3d$ - O $2p$ orbital hybridization (bandwidth) and/or lower the occupied out-of-plane Cr $3d$ $t_{2g}$ (xz and yz) orbitals that would tend to be more delocalized and increase the hopping transport probability (Figure 6). The orbital configurations proposed here in Figure 6 have obviously to be confirmed by further studies. Various ground state configurations are possible from more complex spin-orbital-lattice couplings with different orbital occupancy/ordering and consequent magnetic ordering [23, 55] since the orbital energy levels and orbital occupancy in $CrO_2$ can be easily switched [67]. Additional more complex features may also occur, such as orbital ordering/patterning or bond disproportionation as observed with other $3d$ perovskite oxides [47, 69].

Alternatively, slight cationic stoichiometry deviations or introduction of point/planar defects in the lattice depending on strain, driven by partial strain relaxation, could also significantly impact the transport properties [34, 37, 70]. However, *i*) the films here are epitaxially strained (Fig. 2b), *ii*) the film composition has been checked by XPS to be at the targeted stoichiometry, although the instrumental uncertainty is larger than 5 at.% (Supporting Information, Fig. S6) [11, 27], *iii*) proper flux measurements and composition calibrations have been performed with our best accuracy (less than 3 at.%, Supporting Information, Fig. S1) [27], *iv*) the transport properties of LSCO film on $SrTiO_3$(001) substrate are in perfect agreement with literature and the strain-dependent trend (Supporting Information, Figs. S9-S10) [24], *v*) the *c/a* ratio has a linear trend with in-plane strain, as expected from Poisson ratio at this composition [27], and *vi*) the mosaicity of the films is below 0.1° on LSAT and DSO and around 0.2° on LAO, partially caused by substrate quality (Supporting Information, Fig. S2), meaning a low defect density in the films [27].

Although accurate microscopic physical mechanisms responsible for these strain-dependent transport properties of LSCO would require further investigations, our experimental results present a robust basis and already show the critical impact of strain on the LSCO transport properties, which is of high interest for LSCO-based applications and devices.

## IV. CONCLUSION

In summary, high-quality *p*-type thermoelectric transparent $La_{0.75}Sr_{0.25}CrO_3$ (LSCO) epitaxial thin films have been grown by molecular beam epitaxy on different (pseudo)cubic (001)-oriented perovskite-oxide substrates: $LaAlO_3$, $(LaAlO_3)_{0.3}(Sr_2AlTaO_6)_{0.7}$, and $DyScO_3$, which results in epitaxial strain ranging from -2.06 % to +1.75 %. We have shown that the structural properties (lattice parameters, unit-cell volume, density), as well as electronic transport (conductivity, hole concentration, and mobility) and thermoelectric properties (Seebeck coefficient and power factor), can be efficiently tuned by epitaxial strain. At room temperature, electrical conductivity that is consistent with thermally activated hopping transport increases by two orders of magnitude, from ~0.5 S cm$^{-1}$ (tensile) to ~35 S cm$^{-1}$ (compressive), whereas the Seebeck coefficient increases by almost a factor of two, from 127 µV K$^{-1}$ (compressive) to 208 µV K$^{-1}$ (tensile), coherently. These results show that the thermoelectric power factor increases by one order of magnitude, from ~2 µW m$^{-1}$ K$^{-2}$ (tensile) to ~56 µW m$^{-1}$ K$^{-2}$ (compressive), and that it can be interestingly enhanced by compressive strain by more than a factor of three with respect to the bulk value. Hole density has been estimated to decrease with electrical conductivity from compressive to tensile strain, indicating a certain concomitant decrease of the Cr $3d$ $t_{2g}$ - O $2p$ orbital hybridizations on the valence band. These results are of high interest for engineering various LSCO-based devices such as thermal energy harvesters, coolers, transparent diodes/electrodes, photocatalyzers or spintronic memories.

## ASSOCIATED CONTENT

**Supporting information:**

MBE flux measurements (BEP-T), Structural characterizations by X-ray diffraction and reflectometry (XRD & XRR), Surface characterizations by atomic force microscopy (AFM), Chemical characterizations by X-ray photoemission spectroscopy (XPS), Summary of the structural properties, Electrical characterizations, Thermoelectric characterizations, Summary of the physical properties at room temperature, References of the supporting information.

## AUTHOR INFORMATION

**Corresponding Author**

*Email: romain.bachelet@ec-lyon.fr

**Author Contributions**

R.B. initiated this work and supervised it with G.S.-G. D.H. prepared the samples and performed the structural (RHEED, XRD-XRR), chemical (XPS) characterizations, and some thermoelectric characterizations. R.M. and M.d'E. contributed to the electric and thermoelectric characterizations. I.F. performed the electrical resistivity characterizations in PPMS. V.M.G., R.D., and S.P. contributed to the thermoelectric characterizations. C.B. contributed to XPS characterizations and thermoelectric characterizations at INL. G.G. contributed to XPS characterizations and discussions on band structure and orbital physics. D.H. and R.B. wrote the manuscript with contributions of all authors.

**Notes**

The pseudo-cubic bulk lattice constant ($a_{bulk}$) of La$_{0.75}$Sr$_{0.25}$CrO$_3$ is 3.876 Å (see Ref. 28), whereas those of LAO, LSAT, STO and DSO substrates ($a_{sub}$) are 3.796 Å, 3.868 Å, 3.905 Å and 3.944 Å, respectively. The out-of-plane and in-plane epitaxial strain ($\varepsilon_{zz}$ and $\varepsilon_{xx}$, respectively) due to the lattice mismatch $f = (a_{sub} - a_{film})/a_{film}$ are defined here as: $\varepsilon_{zz} = (c_{film} - a_{bulk})/a_{bulk}$ and $\varepsilon_{xx} = (a_{film} - a_{bulk})/a_{bulk}$, where $c_{film}$ and/or $a_{film}$ is the measured out-of-plane/in-plane lattice parameter of the film, $a_{bulk}$ the bulk lattice constant.

The thermal expansion coefficients at room temperature of LCO, LAO, LSAT, STO, and DSO are $8.5 \times 10^{-6}$ K$^{-1}$, $10 \times 10^{-6}$ K$^{-1}$, $8.2 \times 10^{-6}$ K$^{-1}$, $9 \times 10^{-6}$ K$^{-1}$, and $8.4 \times 10^{-6}$ K$^{-1}$, respectively.

For the stress estimation, the Young modulus has been taken as the one of cubic LaCrO$_3$ along the [001] axis ($E$ = 316.5 GPa). Retrieved from: https://materialsproject.org/materials/mp-18841/.

Parameters of GaAs-based III-V semiconductors and heterostructures, retrieved from: http://www.ioffe.ru/SVA/NSM/Semicond/index.html.

**ACKNOWLEDGMENTS**

Financial support from the European Commission through the project TIPS (H2020-ICT-02-2014-1-644453), the French national research agency (ANR) through the projects MITO (ANR-17-CE05-0018), LILIT (ANR-16-CE24-0022), DIAMWAFEL (ANR-15-CE08-0034-02), the CNRS through the MITI interdisciplinary programs (project NOTE), IDEX Lyon-St-Etienne through the project IPPON, the Spanish Ministerio de Ciencia e Innovación, through the "Severo Ochoa"

Programme for Centres of Excellence in R&D (SEV-2015-0496) and the MAT2017-85232-R (AEI/FEDER, EU), PID2019-107727RB-I00 (AEI/FEDER, EU), and from Generalitat de Catalunya (2017 SGR 1377) is acknowledged. The China Scholarship Council (CSC) is acknowledged for the grant of Dong Han. Ignasi Fina acknowledges Ramón y Cajal contract RYC-2017-22531. Seebeck measurements at ILM were made within the ILMTech transport platform. The authors are also grateful to Jean-Baptiste Goure, Philippe Regreny, Aziz Benamrouche and Bernat Bozzo for their technical support, and the reviewers for their valuable and constructive comments that have improved the quality of the manuscript.

**GRAPHICAL ABSTRACT**

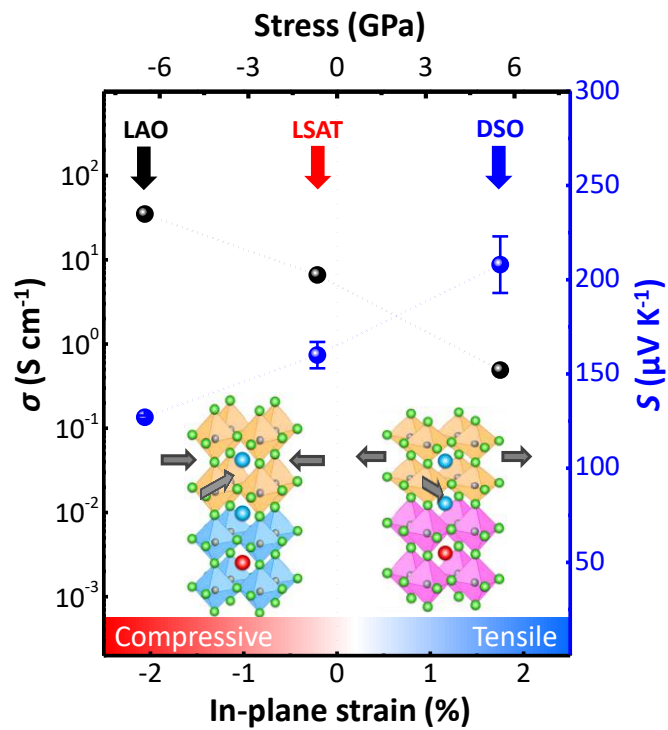